\documentclass[floats,showpacs,aps,twocolumn]{revtex4}
\usepackage{epsfig}
\usepackage{times}
\usepackage{amsmath,amssymb,wasysym}

\newlength\figurewidth
\newcommand{\vc}[1]{\vec{#1}}

\newcommand{\limacon}{{lima\c con}}
\newcommand{\refomega}{\omega_{\mbox{\footnotesize r}}}

\newcommand{\quoting}[1]{``#1''}

\newcommand{\sect}{Sec.}

\newcommand{\rem}[1]{}

\newcommand{\imag}[1]{\text{Im}(#1)}

\newcommand{\real}[1]{\text{Re}(#1)}
\newcommand{\realb}[1]{\text{Re}[#1]}
\newcommand{\imagc}[1]{\text{Im}\,#1}
\newcommand{\realc}[1]{\text{Re}\,#1}

\begin{document}
\title{Electromagnetic modes in cavities made of negative-index metamaterials}
  \author{Jan Wiersig, Julia Unterhinninghofen}
  \affiliation{Institut f{\"u}r Theoretische Physik, Universit{\"a}t Magdeburg,
  Postfach 4120, D-39016 Magdeburg, Germany}
  \author{Henning Schomerus}
  \affiliation{Department of Physics, Lancaster University, Lancaster LA1 4YB, United Kingdom}
  \author{Ulf Peschel}
\affiliation{Institute for Optics, Information and Photonics, University Erlangen-Nuremberg, D-91058, Erlangen, Germany}
  \author{Martina Hentschel}
\affiliation{Max-Planck-Institut f\"ur Physik komplexer Systeme, N{\"o}thnitzer Str. 38, D-01187 Dresden, Germany}
\date{\today}
\begin{abstract}
We discuss electromagnetic modes in cavities formed by metamaterials with negative
refraction and demonstrate that the straightforward approach to substitute negative values of the electric permittivity and the magnetic permeability leads to quasi-bound states with a negative quality factor. To ensure positive quality factors and a consistent physical interpretation of the quasi-bound states it is essential to include the frequency dispersion of the permittivity and the permeability, as required by positive field energy and causality. The basic mode equation and the boundary conditions including linear frequency dispersion are derived.  As an example we consider a disk-like cavity with deformed cross sectional shape. The transition from the unphysical nondispersive case with negative quality factors to the dispersive case with positive quality factors is demonstrated numerically and in an analytical perturbative treatment.
\end{abstract}

\pacs{42.25.-p, 42.60.Da, 78.67.Pt}
\maketitle

\section{Introduction}
Negative-index metamaterials (NIMs) are artificial composites
characterized by simultaneously negative values of the electric permittivity
$\varepsilon$ and the magnetic permeability
$\mu$~\cite{Ramakrishna05,EZ05,HAH07}. These materials were theoretically
predicted already in 1968 by Veselago~\cite{Veselago68}. In present days there is a strong interest in such materials because of potential applications, such as electromagnetic cloaking~\cite{SMJ06}, subwavelength imaging~\cite{Pendry00} and focussing of light~\cite{SHD07}.

Cavities which confine electromagnetic waves in all three spatial dimensions have also attracted considerable attention in the recent years, in particular in the optical regime~\cite{Vahala03}. The strong interest is partly due to the numerous future applications, such as single-photon emitters~\cite{Michler2000} and ultralow threshold lasers~\cite{Park04,SCL08,WGJ09}, and partly due to the possibility to address fundamental questions of light-matter interaction~\cite{RSL04} and ray-wave correspondence~\cite{ND97,GCNNSFSC98,SLK08}.

The fabrication of electromagnetic cavities made of NIMs is rather
challenging with nowadays technology. The literature in this field is
therefore limited to a few theoretical studies, e.g., on
one-dimensional cavities made of distributed Bragg reflectors~\cite{SGL07} and
two-dimensional superscatterers~\cite{YCLM08}. These works, however, consider
scattering of plane waves and do not treat the electromagnetic modes as quasi-bound
states with finite lifetime.
In this paper we fill this gap and discover a subtle difficulty when defining
quasi-bounded states in the canonical way. We show that for a consistent
physical interpretation of modes in NIMs the frequency
dispersion of $\varepsilon$ and $\mu$ is crucial. Our consideration is general 
and applies to all sorts of NIM cavities. For illustration we present results 
for two-dimensional disk cavities with deformed cross-sectional shape.

The paper is organized as follows. Sections~\ref{sec:Maxwell} and ~\ref{sec:conventional} provide a brief review on Maxwell's equations for monochromatic fields and on electromagnetic cavities made of conventional materials. The boundary conditions for NIM cavites and the appearance of negative quality factors are discussed in \sect~\ref{sec:bcNIM}. Section~\ref{sec:dispersion} deals with the basic properties of the frequency dispersion of $\varepsilon$ and $\mu$. A modified mode equation including linear dispersion and a numerical solution of this equation is presented in \sect~\ref{sec:MME}. A discussion of the effects of the dispersion is given in \sect~\ref{sec:discussion}. Finally, \sect~\ref{sec:conclusions} contains the conclusions. 

\section{Maxwell's equations}
\label{sec:Maxwell}
The source-free Maxwell's equations in the frequency domain are
\begin{eqnarray}\label{eq:Maxwell}
\nabla\times\vc{E} =  i\frac{\omega}{c}\vc{B} \, , &&
\nabla\times\vc{H} = -i\frac{\omega}{c}\vc{D}\, , \\
\label{eq:Maxwell2}
\nabla\cdot\vc{D}=0\, , &&
\nabla\cdot\vc{B}=0\, ,
\end{eqnarray}
where $c$ is the speed of light in vacuum.
The constitutive relations for monochromatic fields with frequency
$\omega$ are in the isotropic and linear regime
\begin{eqnarray}\label{eq:constitutive1}
\vc{D}(\vc{r},\omega) & = & \varepsilon(\vc{r},\omega)\vc{E}(\vc{r},\omega)\, ,\\
\label{eq:constitutive2}
\vc{B}(\vc{r},\omega) & = & \mu(\vc{r},\omega)\vc{H}(\vc{r},\omega) 
\end{eqnarray}
with electric permittivity $\varepsilon$ and magnetic permeability $\mu$.
The boundary conditions at an interface between a material 1 and a
material 2 are given by
\begin{eqnarray}\label{eq:bcgeneral1}
\vc{\nu}\times(\vc{E}_1-\vc{E}_2) & = 0 = & \vc{\nu}\times(\vc{H}_1-\vc{H}_2)\, ,\\
\label{eq:bcgeneral2}
\vc{\nu}\cdot(\mu_1\vc{H}_1-\mu_2\vc{H}_2) & = 0 = & \vc{\nu}\cdot(\varepsilon_1\vc{E}_1-\varepsilon_2\vc{E}_2)\, ,
\end{eqnarray}
where $\vc{\nu}$ is the local normal vector on the boundary.

\section{Conventional cavities}
\label{sec:conventional}
When discussing electromagnetic modes in cavities made of conventional materials one usually takes advantage of two
simplifications: (i) the frequency dispersion of the  permittivity $\varepsilon$
 is ignored, assuming that the frequency
interval of interest is sufficiently small. (ii) The permeability $\mu$
is assumed to be constant and unity at all frequencies, throughout the whole space.

In the following, we focus on disk-like cavities mainly for illustration purposes. We emphasise here that our results can be extended to arbitrarily shaped three-dimensional cavities in a straightforward manner. 
For the quasi-2D geometry of the disk one separates the $(x,y)$-dynamics in the plane of the disk from the $z$-dynamics by expanding the electric field in terms of $\vc{E}(x,y)e^{ink_zz}$, with $k_z = 0$ or $k_z$ finite. In the latter case the refractive index $n=\sqrt{\varepsilon\mu}$ in the mode equation can be replaced by an effective index.  Assuming a piecewise constant
index of refraction one can derive the following mode
equation for a quasi-two-dimensional disk~\cite{Jackson83eng,TSSJ05}
\begin{equation}\label{eq:wave}
-\nabla^2\psi = n^2\frac{\Omega^2}{c^2}\psi \ ,
\end{equation}
where $\Omega$ is the frequency of the mode. This Helmholtz equation holds for both transverse
magnetic (TM) and transverse electric (TE) polarization. For TM polarization
the electric field is perpendicular to the cavity plane with $E_z =
\realb{\psi(x,y)e^{-i\Omega t}}$. At the boundary between a material 1 and a
material 2 the general boundary conditions~(\ref{eq:bcgeneral1}) and (\ref{eq:bcgeneral2}) give the following continuity relations for the wave function $\psi$ and
its normal derivative $\partial_\nu\psi$ along the normal $\vc{\nu}$ ,
\begin{equation}\label{eq:bcTM}
\psi_1 = \psi_2 \:,\, \partial_\nu\psi_1 = \partial_\nu\psi_2 \quad\mbox{(TM)},
\end{equation}
assuming that $\mu_1=\mu_2$. For TE polarization, the magnetic field is perpendicular to the cavity plane with $H_z =
\realb{\psi(x,y)e^{-i\Omega t}}$. The boundary conditions are
\begin{equation}\label{eq:bcTE}
\psi_1 = \psi_2 \;,\, \frac{1}{n_1^{2}}\partial_\nu\psi_1 = \frac{1}{n_2^{2}}\partial_\nu\psi_2 \quad\mbox{(TE)},
\end{equation}
again assuming $\mu_1=\mu_2$. At infinity, outgoing wave conditions in the two-dimensional disk plane
\begin{equation}\label{eq:outgoingbc}
\psi \sim \psi_{\text{out}} =
h(\theta,k)\frac{\exp{(ikr)}}{\sqrt{r}}
\end{equation}
with wave number $k=\Omega/c$ are imposed for both polarizations, which results in
quasi-bound states with  frequencies $\Omega$ situated in the lower half of
the complex plane. Whereas the real part is the usual frequency, the
imaginary part is related to the lifetime $\tau=-1/[2\,\imagc{\Omega}]$. The
quality factor of a quasi-bound state is defined by $Q =
-\realc{\Omega}/[2\,\imagc{\Omega}]$.
These resonant states, first introduced by Gamow~\cite{Gamow28} and
by Kapur and Peierls~\cite{KP38}, are connected to the peak
structure in scattering spectra; see~\cite{Landau96} for an introduction.

\begin{figure}[ht]
\centerline{\includegraphics[width=0.6\figurewidth]{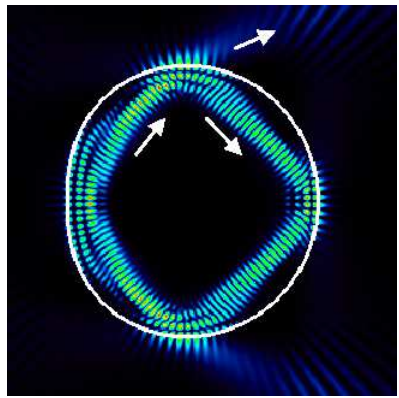}}
\caption{(color online). Near-field intensity pattern of an electromagnetic mode with
normalized frequency $\Omega R/c = 44.9376-i0.0622$ in a conventional cavity with
$\varepsilon = 9/4$ and $\mu = 1$ ($n=1.5$) surrounded by vacuum with
  $\varepsilon = \mu = 1$. Arrows illustrate the direction of the energy flow. }
\label{fig:modeplus}
\end{figure}

As an example we choose a disk-like cavity with the boundary curve being the
{\limacon} of Pascal, which reads in polar coordinates
\begin{equation}
\rho(\phi) = R(1+e\cos\phi) \ .
\end{equation}
For vanishing deformation parameter $e$ this gives the circular disk with
radius~$R$. We choose a deformed disk with $e=0.43$. Exactly this geometry 
has been studied for conventional materials in the context of directed light 
emission from microlasers, theoretically~\cite{WH08} as well as
experimentally~\cite{YWD09,SHH09,YKK09,SCL08}. The value of $R$ itself is not relevant, only the ratio $R/\lambda$ is important, where $\lambda=2\pi/k$ is the wavelength. Therefore, we consider in the following a  normalized frequency $\Omega R/c = kR$.
Figure~\ref{fig:modeplus} shows a typical TM polarized mode in a {\limacon} cavity with low index of refraction, $n=1.5$, computed with the boundary element method~\cite{Wiersig02b}. The mode is
localized along an unstable periodic ray trajectory, i.e.,
it is a so-called scarred mode~\cite{Heller84}. From the relative intensity of the different segments we
can assess the direction of energy flow as indicated by the arrows.
The frequency is $\Omega R/c = 44.9376-i0.0622$, the $Q$-factor is therefore about $361$. Such a medium-$Q$, scarred mode is well suited to
demonstrate how the light is
(partially) refracted out. This will be useful in the following comparison to
the NIM cavities.

\section{NIM boundary conditions}
\label{sec:bcNIM}
In the case of a NIM cavity the permeability $\mu$ can no longer to be 
treated as spatially uniform as this quantity changes sign at the interface between of the NIM cavity and the surrounding conventional material (in our case vacuum). In this situation the general
boundary conditions~(\ref{eq:bcgeneral1}) and (\ref{eq:bcgeneral2}) give
(see, e.g., Ref.~\cite{SGL07})
\begin{equation}\label{eq:bcTMMeta}
\psi_1 = \psi_2 \:,\, \frac{1}{\mu_1}\partial_\nu\psi_1 = \frac{1}{\mu_2}\partial_\nu\psi_2 \quad\mbox{(TM)}
\end{equation}
instead of the special case in Eq.~(\ref{eq:bcTM}).
For TE polarization we have
\begin{equation}\label{eq:bcTEMeta}
\psi_1 = \psi_2 \:,\, \frac{1}{\varepsilon_1}\partial_\nu\psi_1 = \frac{1}{\varepsilon_2}\partial_\nu\psi_2 \quad\mbox{(TE)}
\ .
\end{equation}

Using these boundary conditions together with 
the outgoing-wave conditions~(\ref{eq:outgoingbc}) at infinity we find that 
solutions of the mode equation~(\ref{eq:wave}) always have a negative 
$Q$-factor for NIMs, i.e., the intensity of such a solution $\psi$ does not decay but 
instead increases exponentially  in time, which is unphysical for a passive material. As a typical 
example which will be discussed later in more detail we mention a mode in the 
{\limacon} cavity with $\varepsilon = -9/4$ and $\mu = -1$ ($|n|=1.5$). The 
normalized frequency is $\Omega_0 R/c = 45.1060+i0.1647$. The quality factor 
is therefore about $Q_0 = -137$.

Note that for a conventional material a similar effect may occur if a thin 
active layer with very strong gain is placed at the interface. In that case
the jump of the derivative imposed for negative index materials in 
Eq.~(\ref{eq:bcTMMeta}) or (\ref{eq:bcTEMeta}) would be caused by an outflow of energy from that amplifying layer. 

It seems that for a passive NIM the exponential increase of the electromagnetic intensity contradicts the law of energy conservation. This is, however, not the case as can be seen by considering the electromagnetic field energy
density
\begin{equation}\label{eq:W}
W = \frac{1}{8\pi}\left(\varepsilon\vc{E}^2+\mu\vc{H}^2\right) \ .
\end{equation}
In this equation and also in the following ones we suppress the dependency on
the spatial coordinates for notational convenience.
As already pointed out by Veselago~\cite{Veselago68}, as $\varepsilon, \mu < 0$ the field energy~(\ref{eq:W})  in a NIM would be negative and unbounded from below. In our case of a quasi-bound cavity mode this results in an exponential decay of the field energy towards $-\infty$, as the cavity permanently looses
(positive) energy to the outside. Hence, the field intensity $\vc{E}^2>0$ inside the cavity increases exponentially in time. The imaginary part of the frequency $\Omega$ is therefore
positive, and the quality factor negative, consistent with our numerical 
finding.

It is worth mentioning that Dirac's wave equation for relativistic electrons possesses a similar \quoting{radiation catastrophe} which disappears in a proper quantum field theoretical treatment.
For the NIM materials, the problem of negative field energy density can be, however, solved already on the wave equation level, namely by the inclusion of the frequency dispersion of the electric permittivity and the magnetic permeability~\cite{Veselago68,LanLif60eng}. 

\section{Frequency dispersion}
\label{sec:dispersion}
When the dispersion of $\varepsilon$ and $\mu$ is important the expression for the field energy density~(\ref{eq:W}) has to be replaced by
\begin{equation}\label{eq:Wdis}
W =
\frac{1}{8\pi}\left(\frac{\partial(\varepsilon\omega)}{\partial\omega}\vc{E}^2+\frac{\partial(\mu\omega)}{\partial\omega}\vc{H}^2\right)
\ .
\end{equation}
The energy density $W$ defined by Eq.~(\ref{eq:Wdis}) is positive provided that
\begin{equation}
\frac{\partial(\varepsilon\omega)}{\partial\omega} > 0 \;,\quad
\frac{\partial(\mu\omega)}{\partial\omega} > 0
\end{equation}
for any values of $\vc{E}^2$ and $\vc{H}^2$.
These inequalities imply lower bounds for the derivatives
\begin{equation}\label{eq:inequal}
\frac{\partial\varepsilon}{\partial\omega} > -\frac{\varepsilon}{\omega} \;,\quad
\frac{\partial\mu}{\partial\omega} > -\frac{\mu}{\omega} \ ,
\end{equation}
with $\omega > 0$. 
For conventional materials with positive $\varepsilon$ and $\mu$ and nonresonant response to external fields the derivatives are small, and therefore often frequency dispersion can be safely ignored. For NIMs with negative $\varepsilon$ and $\mu$ this is never possible.

In a transparency region, where we can neglect absorption in the material, causality requires the following additional inequalities~\cite{LanLif60eng}
\begin{equation}\label{eq:inequalstrong}
\frac{\partial\varepsilon}{\partial\omega} > \frac{2(1-\varepsilon)}{\omega} \;,\quad
\frac{\partial\mu}{\partial\omega} > \frac{2(1-\mu)}{\omega} \ .
\end{equation}
For negative $\varepsilon$ and $\mu$ these inequalities are stronger than the ones in Eq.(\ref{eq:inequal}).
Note that there is a controversy in the NIM community concerning the existence of negative refraction in such a transparency region; see, e.g., Refs.~\cite{Stockmann07,NS08,KM08}. This controversy is, however, not settled yet; so in the following we assume that a transparency region with negative $\varepsilon$ and $\mu$ exists.

It is worth mentioning that in (unrealistic) materials having no absorption at all in the whole frequency interval, the inequalities in  Eq.~(\ref{eq:inequalstrong}) can turn into equalities. This is, for instance, the case for the nonlossy Drude model with the dielectric function
\begin{equation}
\varepsilon(\omega) = 1-\frac{\omega_p^2}{\omega^2}
\end{equation}
and plasma frequency $\omega_p$. Another, more general model which is often used to describe NIMs locally in frequency space is the Drude-Lorentz system
\begin{equation}
\varepsilon(\omega) = 1-\frac{\Omega^2}{\omega^2-\omega_0^2+i\gamma\omega} \ .
\end{equation}
In the limiting case of no absorption $\gamma\to0$ the  inequalities in  Eq.~(\ref{eq:inequalstrong}) are fulfilled. For the subtle issue of causality in the nonlossy Drude-Lorentz model we refer the reader to Ref.~\cite{Tip04}.

\section{Modified mode equation}
\label{sec:MME}
In the following, we show that when dealing with quasi-bound states not only Maxwell's equations but also the constitutive relations including the frequency dispersion have to be extended to the complex frequency plane. This fact directly leads to a modified mode equation which is capable of describing linear frequency dispersion. This mode equation predicts positive quality factors in agreement with the requirements of positive field energy and causality.

The key observation to start with is that a quasi-bound state
\begin{equation}\label{eq:mode}
\vc{E}_\Omega(t) = \vc{E}_0 e^{-i\Omega t}
\end{equation}
with complex-valued frequency $\Omega$ is not a monochromatic wave as its Fourier decomposition  
\begin{equation}
\vc{E}_\Omega(t) = \int_{-\infty}^\infty d\omega\,
\vc{E}(\omega)e^{-i\omega t} 
\end{equation}
gives nonvanishing $\vc{E}(\omega)$ in a region around $\omega\approx\real{\Omega}$. With this decomposition we can write the constitutive
relation~(\ref{eq:constitutive1}) as
\begin{equation}\label{eq:D1}
\vc{D}_\Omega(t) = \int_{-\infty}^\infty d\omega\,
\varepsilon(\omega)\vc{E}(\omega)e^{-i\omega t} \ .
\end{equation}
To proceed further, let us fix a real-valued frequency $\refomega$ around which we study electromagnetic
modes. We restrict ourselves to modes with sufficiently small $|\imag{\Omega}|$,
i.e., not too small $Q$-factor, and $\real{\Omega} \approx \refomega$. This
allows the expansion of the permittivity
\begin{equation}\label{eq:expand}
\varepsilon(\omega) \approx
\varepsilon(\refomega)+\frac{\partial\varepsilon}{\partial\omega}\Big|_{\refomega}(\omega-\refomega) \ .
\end{equation}
Inserting this expansion into Eq.~(\ref{eq:D1}) gives
\begin{eqnarray}
\vc{D}_\Omega(t) & = & \varepsilon(\refomega)\vc{E}_\Omega(t)
+\frac{\partial\varepsilon}{\partial\omega}\Big|_{\refomega}\left(i\frac{\partial}{\partial t}-\refomega\right) \vc{E}_\Omega(t) .
\end{eqnarray}
Now we exploit the property of quasi-bound states $\frac{\partial}{\partial
  t}\vc{E}_\Omega = -i\Omega\vc{E}_\Omega$, which can be deduced from Eq.~(\ref{eq:mode}) and leads to
\begin{eqnarray}\label{eq:analyticcontinuation}
\vc{D}_\Omega(t) & = & \tilde\varepsilon(\Omega)\vc{E}_\Omega(t)\, ,
\end{eqnarray}
with  modified permittivity
\begin{equation}\label{eq:effe0}
\tilde\varepsilon(\Omega) =
\varepsilon(\refomega)+\frac{\partial\varepsilon}{\partial\omega}\Big|_{\refomega}\left(\Omega-\refomega\right)
\ .
\end{equation}
Equation~(\ref{eq:analyticcontinuation}) represents an analytic continuation of the permittivity $\varepsilon(\omega)$ to the complex-frequency plane $\Omega$. For this continuation the linearization in Eq.~(\ref{eq:expand}) is not needed. In fact, for the modified mode equation that we derive in the following, the extension to any analytic function $\varepsilon(\omega)$ is straightforward. Nevertheless, for clarity we restrict ourself to a linear frequency dispersion. 
We would like to point out that the imaginary part of $\tilde\varepsilon$ is related to the frequency dispersion and not to optical absorption in the material, which is neglected here. Nevertheless, a realistic $\Omega\neq\refomega$ with negative imaginary part introduces a kind of loss in the originally lossless medium which counteracts the nonphysical exponential increase of the intensity, thereby turning the negative quality factor into a positive one.

In the following it will be convenient to express the values of the
derivatives of $\varepsilon$ and $\mu$ at the fixed frequency $\refomega$
by their dimensionless linear dispersions
\begin{equation}\label{eq:dimless}
\alpha_\varepsilon =
-\frac{\partial\varepsilon}{\partial\omega}\Big|_{\refomega}
\frac{\refomega}{\varepsilon} \;,\quad
\alpha_\mu = -\frac{\partial\mu}{\partial\omega}\Big|_{\refomega} \frac{\refomega}{\mu} \ .
\end{equation}
For NIMs ($\varepsilon<0$, $\mu<0$) these quantities $\alpha_\varepsilon$,
$\alpha_\mu$ have to be chosen larger than 1 to satisfy the
inequalities~(\ref{eq:inequal}) (for conventional dielectrics $\alpha_\varepsilon$,
$\alpha_\mu$ must be smaller than 1.). To satisfy the inequalities~(\ref{eq:inequalstrong}) for NIMs we must have
\begin{equation}\label{eq:constrains}
\alpha_\varepsilon > 2-\frac{2}{\varepsilon}\;,\quad
\alpha_\mu > 2-\frac{2}{\mu} \ .
\end{equation}
With the quantities in Eq.~(\ref{eq:dimless}) we can rewrite Eq.~(\ref{eq:effe0}) as
\begin{equation}\label{eq:effe}
\tilde \varepsilon(\Omega) =
\varepsilon(\refomega)\left(1+\alpha_\varepsilon\frac{\refomega-\Omega}{\refomega}\right) \ .
\end{equation}
We can derive a modified permeability in an analogue way
\begin{equation}\label{eq:effm}
\tilde \mu(\Omega) = \mu(\refomega)\left(1+\alpha_\mu\,\frac{\refomega-\Omega}{\refomega}\right)
\ .
\end{equation}
As a result of our considerations we can use Maxwell's equations~(\ref{eq:Maxwell})-(\ref{eq:Maxwell2}) and the
constitutive relations~(\ref{eq:constitutive1})-(\ref{eq:constitutive2}) for monochromatic
waves with modified permittivity and permeability given by
Eqs.~(\ref{eq:effe}) and (\ref{eq:effm}). A direct consequence is the modified
mode equation
\begin{equation}\label{eq:waveselfconsistent}
-\nabla^2\psi =
\tilde{n}^2(\Omega)\frac{\Omega^2}{c^2}\psi \ ,
\end{equation}
with the modified refractive index
\begin{eqnarray}\label{eq:effn}
\tilde n(\Omega) & = & \sqrt{\tilde\varepsilon(\Omega)\tilde\mu(\Omega)}
\approx n(\refomega)\left(1+\alpha_n\,\frac{\refomega-\Omega}{\refomega}\right)
\end{eqnarray}
and the dimensionless linear dispersion
\begin{equation}\label{eq:an}
\alpha_n=-\frac{\partial n}{\partial \omega}\Big|_{\refomega}\frac{\refomega}{n} = \frac{\alpha_\varepsilon+\alpha_\mu}{2} \ .
\end{equation}
In the derivation we have ignored terms of order $(\Omega-\refomega)^2$, which 
is consistent with Eq.~(\ref{eq:expand}). For the square root in 
Eq.~(\ref{eq:effn}) we choose the positive branch. Note that the (modified) 
refractive index can be defined negative or positive. This does not matter for 
our purpose, as the sign of the refractive index neither enters the
mode equation~(\ref{eq:waveselfconsistent}) nor the modified boundary
conditions
\begin{eqnarray}
\label{eq:modbcTMMeta}
\psi_1 & = & \psi_2 \:,\, \frac{1}{\tilde\mu_1}\partial_\nu\psi_1 = \frac{1}{\tilde\mu_2}\partial_\nu\psi_2 \quad\mbox{(TM)}\\
\label{eq:modbcTEMeta}
\psi_1 & = & \psi_2 \:,\, \frac{1}{\tilde\varepsilon_1}\partial_\nu\psi_1 = \frac{1}{\tilde\varepsilon_2}\partial_\nu\psi_2 \quad\mbox{(TE)}
\ .
\end{eqnarray}
The phenomenon of negative refraction is here a result of the relative sign of the permittivity and the permeability in the boundary conditions~(\ref{eq:modbcTMMeta}) and (\ref{eq:modbcTEMeta}).

After the quantities $\varepsilon$, $\mu$ and their first derivative are
specified at a given reference frequency $\refomega$, the modified mode
equation~(\ref{eq:waveselfconsistent}), the modified
permittivity~(\ref{eq:effe}), the modified permeability~(\ref{eq:effm}), the 
modified refractive index~(\ref{eq:effn}), and the modified boundary
conditions~(\ref{eq:modbcTMMeta})-(\ref{eq:modbcTEMeta}) have to be solved self-consistently.
This can be done with only slight modifications of standard approaches, such
as the boundary element method~\cite{Wiersig02b}.

\begin{figure}[ht]
\centerline{\includegraphics[width=1.0\figurewidth]{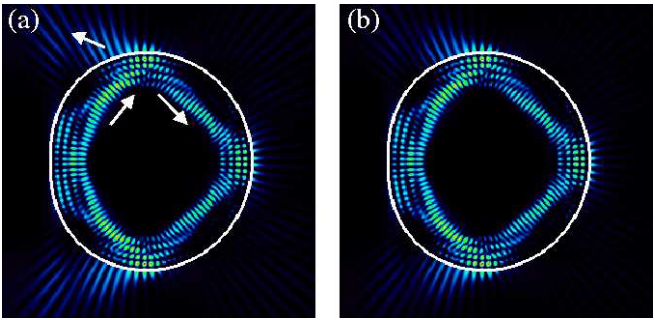}}
\caption{(color online). (a) Near-field intensity pattern of an electromagnetic mode
  with $\Omega R/c = 45.0966-i0.0514$ in a NIM cavity with
  $\varepsilon(\refomega) = -9/4$, $\mu(\refomega) = -1$ ($|n(\refomega)|=1.5$), $\refomega R/c = 45.1$, and linear frequency dispersion
  $\alpha_\varepsilon=\alpha_\mu=4.1$. Arrows illustrate the direction of the energy
  flow. (b) Unphysical mode with $\Omega_0 R/c = 45.1060+i0.1647$ in a NIM cavity with
  $\varepsilon = -9/4$ and $\mu = -1$
  ($|n|=1.5$) calculated without frequency dispersion.
}
\label{fig:modeminus}
\end{figure}

As an example we consider again a mode in the {\limacon} cavity with 
normalized frequency $\real{\Omega}R/c$ around $45.1$. We fix the reference 
frequency $\refomega R/c$ therefore to be $45.1$. Note that the precise value of the reference frequency is not relevant as long as we choose the values for the  permittivity, the permeability and their derivatives accordingly, e.g. by reading off their values from a material dispersion curve at the given reference frequency. For the NIM we consider $\varepsilon(\refomega) = -9/4$ and $\mu(\refomega) = -1$ as in the previous section. To fulfill the constraints from Eq.~(\ref{eq:constrains})
\begin{equation}
\alpha_\varepsilon > 2+\frac{8}{9}\;,\quad
\alpha_\mu > 4 \ ,
\end{equation}
we first set $\alpha_\varepsilon = \alpha_\mu = 4.1$ and vary this value later. According to Eq.~(\ref{eq:an}) the quantity $\alpha_n$ is then also $4.1$.
Figure~\ref{fig:modeminus}(a) shows as an example a TM polarized mode. The $Q$-factor turns out to be positive, $Q = 438$, due to the inclusion of the dispersion.
Interestingly, the spatial mode pattern is rather insensitive to the inclusion of frequency dispersion, as a closer comparison of the dispersive case in Fig.~\ref{fig:modeminus}(a) and the (unphysical)
nondispersive case in Fig.~\ref{fig:modeminus}(b) shows.

\begin{figure}[ht]
\centerline{\includegraphics[width=\figurewidth]{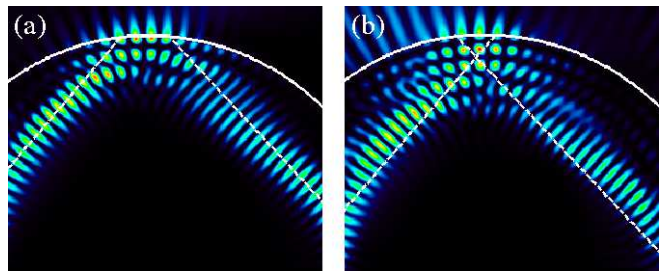}}
\caption{(color online). Goos-H\"anchen shift in (a) a conventional cavity (cf.\ Fig.~\ref{fig:modeplus}), 
and (b) a NIM cavity [cf.\ Fig.~\ref{fig:modeminus}(a)]. The dotted line is the center of a Gaussian beam being reflected at the dielectric interface. The appearance of the Goos-H\"anchen shift along the boundary is evident. The mode pattern is the full numerical solution of the respective mode equation.}
\label{fig:ghs}
\end{figure}

Contrasting the mode in the NIM cavity in Fig.~\ref{fig:modeminus}(a) with a
corresponding mode in a conventional material in Fig.~\ref{fig:modeplus}
clearly reveals the negative refraction. To be more precise, one part of the
beam is confined by total internal reflection and another part is refracted out.
Moreover, a careful inspection shows a different Goos-H\"anchen shift (GHS)
for the NIM and for the conventional material, as can be seen in
Fig.~\ref{fig:ghs}. The GHS is a lateral shift of totally reflected beams
along the optical interface due to interference~\cite{GH47} (for GHS in
cavities see Refs.~\cite{HS06,UWH08}).  According to Ref.~\cite{Berman02}, a
light beam in a conventional material reflected at the interface to a NIM
experiences a negative GHS. In our case the light propagates in the NIM and
is reflected at the interface to a conventional material. We compute the GHS
by reflecting a Gaussian beam at a planar dielectric interface, neglecting
boundary curvature effects \cite{HS02}. For the mode in the conventional cavity, see
Fig.~\ref{fig:ghs}(a), we find a nearly perfect agreement between such a beam
reflection at a planar interface and the full mode calculation. For the mode
in the NIM cavity, see Fig.~\ref{fig:ghs}(b), the agreement is somewhat 
reduced, but
nevertheless the appearance of a negative GHS can be clearly seen.

\section{Discussion}
\label{sec:discussion}
\begin{figure}[ht]
\centerline{\includegraphics[width=1.0\figurewidth]{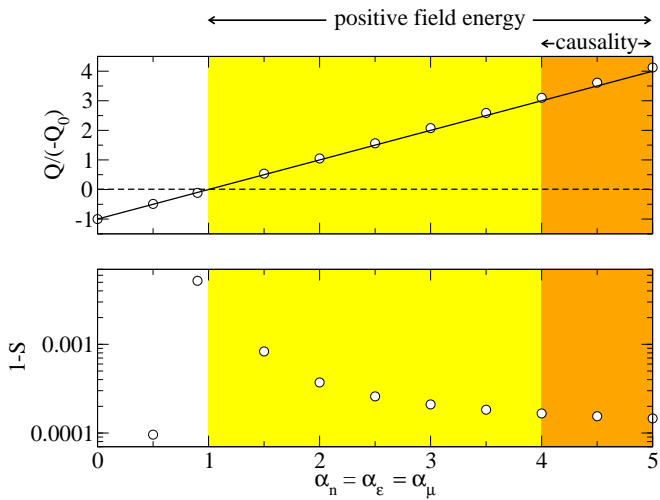}}
\caption{(color online). $Q/(-Q_0)$ (upper panel) and overlap difference $1-S$ [lower panel,
  cf.\ Eq.(\ref{eq:overlap})] vs.\ linear frequency dispersion $\alpha_n=\alpha_\varepsilon = \alpha_\mu$. All quantities are dimensionless. Here, $Q_0=-137$ is the
  quality factor in the nondispersive case $\alpha_n=0$. In the regime $\alpha_n>1$ the electromagnetic field energy is positive and for $\alpha_n>4$ also causality holds. The empty dots mark the results of the full solution of the mode equation~(\ref{eq:waveselfconsistent}). The solid line is the theoretical prediction in Eq.~(\ref{eq:QQ0}).}
\label{fig:theory}
\end{figure}

To quantify small differences in the spatial pattern of the mode $\psi_0$ without
frequency dispersion [as, e.g., in Fig.~\ref{fig:modeminus}(b)] and a mode $\psi$
with dispersion [as, e.g., in Fig.~\ref{fig:modeminus}(a)] we examine the normalized spatial overlap
\begin{equation}
\label{eq:overlap}
S = \frac{|\int_{\cal C} dxdy\;\psi_0^*\psi|}{\sqrt{\int_{\cal C} dxdy\;\psi_0^*\psi_0}\sqrt{\int_{\cal C} dxdy\;\psi^*\psi}} \ .
\end{equation}
We restrict the integrals to the interior of the cavity~${\cal C}$ as the exterior is influenced by the actual value of the quality factor.
For the modes in  Fig.~\ref{fig:modeminus} (a) and (b) we find $1-S \approx 1.6\times 10^{-4}$, so indeed the overlap is nearly unity. The lower panel of Fig.~\ref{fig:theory} shows $1-S$ as function of the linear frequency dispersion~$\alpha_n$. Except near $\alpha_n = 1$, the overlap $1-S$ is below $0.001$. The upper panel of Fig.~\ref{fig:theory} shows $Q/(-Q_0)$ with $Q_0 = -137$ as function of $\alpha_n$. It can be observed that for $\alpha_n>1$, where the field energy is positive,  the $Q$-factor is also positive.

Can we understand the basic features observed in Fig.~\ref{fig:theory}? To do
so, let us first note that the quantities $|(\tilde n-n)/n|$,
$|(\tilde\varepsilon-\varepsilon)/\varepsilon|$, and $|(\tilde\mu-\mu)/\mu|$ are
small, as required by the linear expansions, e.g., in Eq.~(\ref{eq:expand}). The applicability of perturbation theory to the Helmholtz equations~(\ref{eq:wave}) and (\ref{eq:waveselfconsistent}) therefore implies that the spatial mode pattern does not depend much on the frequency dispersion. But why is the $Q$-factor so strongly dependent on the frequency dispersion? This can be understood from the observation that for long-lived modes with high $Q =-\realc{\Omega}/[2\,\imagc{\Omega}]$, the imaginary part of the frequency $\Omega$ is small compared to the real part of $\Omega$. Therefore, even a small modification of $\imag{\Omega}$ in absolute numbers can have a relatively large effect on the quality factor.

To see how $Q$ changes with the linear dispersion $\alpha_n$, consider the mode
equation~(\ref{eq:wave}) with frequency $\Omega_0$ and the modified mode
equation~(\ref{eq:waveselfconsistent}) with frequency $\Omega$. In both cases use the same negative $\varepsilon$, $\mu$. Both mode equations can give the same spatial mode pattern provided that ${n\Omega_0} = {\tilde{n}\Omega}$. From this relation we find  for $|\refomega-\real{\Omega}| \ll \refomega$ and $|\imag{\Omega}|\ll \real{\Omega}$ that
\begin{equation}\label{eq:QQ0}
\frac{Q}{-Q_0} \approx \alpha_n-1 \ .
\end{equation}
This shows that the modified refractive index turns the (non-physical)
negative quality factor $Q_0$ into a positive one, $Q>0$ as soon as $\alpha_n > 1$, i.e., exactly under the condition which 
ensures a positive field energy in the NIM.
To illustrate this relation from another point of view, let us rewrite the right hand side of Eq.~(\ref{eq:QQ0}) as
\begin{equation}\label{eq:velocities}
\alpha_n-1 = -\frac{v_p}{v_g}\, ,
\end{equation}
where $v_p=c/n$ is the phase velocity and $v_g = \partial\omega/\partial k$ is
the group velocity with wave number $k=n\omega/c$. The inequalities~(\ref{eq:inequal}) required by a positive field energy ensure that the phase velocity $v_p$ and the group velocity $v_g$ have a different sign in a NIM. The inequalities~(\ref{eq:inequalstrong}) required by causality carry over to $|v_g| < |v_p|$, i.e., superluminal energy propagation is forbidden.

As demonstrated in Fig.~\ref{fig:theory}, the expression in Eq.~(\ref{eq:QQ0}) is in excellent agreement with the full solution of the modified mode equation~(\ref{eq:waveselfconsistent}).
Note that near $\alpha_n=1$, where $Q\approx 0$, the absolute value of the imaginary part of $\Omega$ and the quantities $|(\tilde n-n)/n|$, $|(\tilde\varepsilon-\varepsilon)/\varepsilon|$, and $|(\tilde\mu-\mu)/\mu|$ are not small. Hence, the linear approximations in our theory are not justified, which explains why in this region $1-S$ is larger.

From another point of view, the expression in Eq.~(\ref{eq:QQ0}) can also be understood purely in real frequency space. Consider a resonant
structure of some spectrum, let's say the Wigner delay time or a scattering cross section. Note that such resonances are determined only by the value of the product $\zeta = n(\omega)\omega$ of refractive index $n$ and the frequency
$\omega$. In our example of the {\limacon} cavity it is $nkR$ rather than just~$kR$, with wave number $k=\omega/c$. Let the full width at half maximum (FWHM) of the resonant peak be $\delta\zeta = n\delta\omega+\omega\delta n$. Comparing the nondispersive case ($\delta n=0$) with the dispersive case we get
\begin{equation}
n\delta\omega_0 = n\delta\omega+\omega\delta n\, ,
\end{equation}
again assuming that the change in the refractive index does not change the spatial mode structure.
With $Q=\omega/\delta\omega$ and $Q_0=\omega/\delta\omega_0$ we arrive after a
few algebraic manipulations at Eq.~(\ref{eq:QQ0}).  

A related expression as in Eq.~(\ref{eq:QQ0}) exists also for conventional materials where it was used to predict enhancement of quality factors in microcavities using highly dispersive materials~\cite{SLH05}. For small group velocity $v_g$, corresponding to slow light, the effect of the $Q$-factor enhancement in Eqs.~(\ref{eq:QQ0}) and (\ref{eq:velocities}) is strongest.
This prediction has been confirmed recently in experiments on slow light in photonic crystals~\cite{GLD08}.

Finally, we note that scattering of a monochromatic wave with (real-valued) frequency $\omega$ at an obstacle made of a NIM is described by the ordinary mode equation~(\ref{eq:wave}) using incoming and outgoing wave conditions \cite{SGL07,YCLM08}. However, when calculating  spectra it is necessary to use the frequency-dependent permittivity $\varepsilon(\omega)$ and permeability $\mu(\omega)$ to be consistent with the requirement of causality. This is usually ignored, perhaps because in a spectrum the FWHM $\delta \omega$ cannot be easily distinguished  from~$-\delta \omega$.

\section{Conclusions}
\label{sec:conclusions}
We addressed quasi-bound electromagnetic modes in negative-index metamaterial cavities.
The simple approach which substitutes negative values of the electric permittivity and magnetic
permeability into the boundary
conditions (which suffices to obtain negative refraction) gives rise to modes with negative
$Q$-factor. This unphysical behavior can be removed by including linear
frequency dispersion in the mode
equation and in the boundary conditions, as is required by a positive field energy and causality.
At complex resonance frequency, the effective permittivity then acquires a finite imaginary part,
which attenuates the mode even in absence of physical absorption.

As an example we studied a disk-like cavity with noncircular cross-sectional shape. The modified mode equation results in modes with positive quality factor and clear signatures of negative refraction and negative Goos-H\"anchen shift.

\begin{acknowledgments}
Financial support from the DFG research group 760, DFG Emmy Noether Programme,
and the European Commission via the Marie Curie Excellence grant
MEXT-2003-02778 is acknowledged.
\end{acknowledgments}


\end{document}